\documentstyle[aas2pp4]{article}

\def\Msun{\ifmmode M_{\odot} \else $M_{\odot}$\fi}
\def\Lsun{\ifmmode L_{\odot} \else $L_{\odot}$\fi}
\def\eg{{\it e.g.,\ }}
\def\ie{{\it i.e.,\ }}
\def\etal{{et al.~}}

\lefthead{Chatzichristou}
\righthead{IR-Warm Seyfert Galaxies}

\begin{document}

\title{Multicolour Optical Imaging of IR-Warm Seyfert Galaxies.\\
    I. Introduction and Sample Selection}

\author{Eleni T. Chatzichristou}
\affil{Leiden Observatory, P.O. Box 9513, 2300 RA Leiden, The Netherlands}

\affil{NASA/Goddard Space Flight Center, Code 681, Greenbelt, MD 20771}



\begin{abstract}
The standard AGN unification models attempt to explain the diversity of 
observed AGN types by a few fundamental parameters, where orientation effects
play a paramount role. Whether other factors, such as the evolutionary stage
and the host galaxy properties are equally important parameters for the AGN
diversity, is a key issue that we are addressing with the present data.
Our sample of IR-selected Seyfert galaxies is based on the important discovery that 
their integrated IR spectrum contains an AGN signature. This being an almost 
isotropic property, our sample is much less affected by orientation/obscuration 
effects compared to most Seyfert samples. It therefore provides a test-bed for the
orientation-dependent models of Seyferts, involving dusty tori.

We have obtained multi-colour broad and narrow band imaging for
a sample of mid-IR ``warm'' Seyferts and for a control sample of mid-IR ``cold'' 
galaxies.  In the present paper we describe the sample selection 
and briefly discuss their IR properties. We then give an overview of the data
collected and present broad-band images for all our objects.
Finally, we summarize the main issues that will be addressed with these data
in a series of forthcoming papers.

\end{abstract}




%

\section{Introduction}

The identification of a new class of objects that emit the bulk of their energy
at infrared wavelengths was one of the most important discoveries of the
IRAS satellite. IR-luminous galaxies become the dominant population of 
extragalactic objects for luminosities greater than
10$^{11}\Lsun$. Their energy output appears to originate from a powerful 
AGN or unusually intense starbursts. There have been arguments suggesting
that all massive galaxies will pass through such an IR-luminous phase 
during their lifetime (\cite{soifer1}).  
Because of the existence of IR-luminous galaxies, statistics based on early 
IRAS results effectively implied that the true space density of AGNs is about
twice the previously supposed value, the majority of the IR selected AGNs being
Seyfert 2s and LINERS (\eg \cite{degrijp3,sanders3}). In the light of this 
discovery, previous results concerning the relative distribution of properties 
among various types of ``active'' galaxies, that were based mainly on optical 
observations, might need to be revised, the IR-luminous galaxies being mostly
too faint to be included in optically selected samples.

The present study represents a systematic investigation of the optical properties of
IR-``warm'' Seyfert galaxies. Their selection is based on the important discovery 
that
their integrated IR spectrum contains an AGN signature (\cite{degrijp1}).
The first intriguing results on these objects (\eg \cite{degrijp2,degrijp3,keel94} motivated the present work. Detailed 
study of their optical properties could provide information about how the AGN is 
triggered (in particular what is the impact of interactions and mergers) and how
the nuclear activity can affect the host galaxy properties. 

A generally accepted 
unification model for AGN postulates that many of their observed characteristics
depend upon the orientation of the observer relative to the dusty torus axis that is
surrounding the central black hole (\eg \cite{antonucci85,antonucci93}). Although
orientation effects are certainly important, of relevance is also the question of
whether an evolutionary sequence exists between starburst and Seyfert activity
(\eg \cite{heckman97,gonzalez98,levenson99}). Recent observational work indeed
suggests that compact nuclear starbursts in Seyfert 2 galaxies play an important 
role, being responsible for their continuum emission (from UV to IR wavelengths) and
significantly contributing to the overall energetics of these objects (\eg 
\cite{maiolino95,heckman95,hunt97,gonzalez98}). Furthermore, it is now recognized
that the intrinsic properties of the massive black hole (mass, accretion, spin)
are also important in the appearance of an AGN. Thus, the universality of the 
orientation unification scheme is far from being proven.

For this kind of studies to be successful and conclusive, the sample definition is
critical so as to rule out any possible selection effects. The mid-IR colour 
selection criterion applied in our sample (i) is based on an unbiased, isotropic
property, (ii) systematically selects objects with high bolometric luminosities, 
whose nuclear IR properties are less affected by the host galaxy, (iii) most 
probably probes (re)radiation from the obscuring torus. This sample is thus
particularly suitable for studying the importance of orientation, as compared to 
other effects, between Seyfert 1 and 2 classes.

Using ground-based multicolour optical imaging data, we have conducted an extensive
study of the optical properties of a mid-IR ``warm'' sample of Seyferts and a
control sample of mid-IR ``cold'' galaxies. In what follows and in future papers these samples will be refered to as Warm and Cold, respectively.
The results will be presented in a 
series of papers. The present introductory paper (Paper I) is structured as follows:
in Section 2 we describe the selection of our subsamples and in Section 3 
we give an overview of the observations and data reduction.
Broad-band contour maps fpr all our objects are presented at the end of this
paper. Finally, in Section 4 we summarize the main issues that we will address
with these data in four forthcoming papers.

\section{Sample Selection}

\begin{figure}
\epsscale{1.1}
\plotone{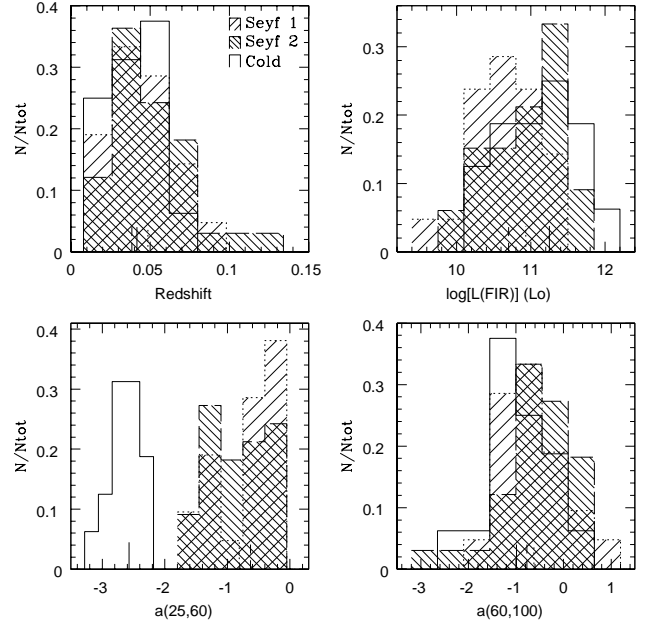}
\caption{The distribution of redshifts, 60-100 $\mu$m luminosities and (25-60) and (60-100) colour indices for the Warm Seyfert 1 and 2 and the Cold samples in this study. The median values are indicated with vertical bars on the lower x-axes: dashed for Seyfert 1s, dotted for Seyfert 2s and solid for Cold galaxies. \label{f2}}
   \end{figure}

\cite{degrijp1} have used the 25-60 $\mu$m colour index as an 
indicator of nuclear activity, the 60 $\mu$m flux being affected 
by the cold galactic dust while the 25 $\mu$m flux is dominated by the nuclear 
component(s). Seyfert galaxies tend to have flatter spectra (``warmer'' colours)
 and represent $\geq$ 70\% of their sample.
During the last few years, we have conducted an observational program to study the 
optical properties of a subsample of Warm Seyferts from the De Grijp \etal
sample. Our objects were selected to cover a large range of IR luminosities over
a relatively narrow redshift range. Here we will present data for 54 of these 
objects, 21 Seyfert 1s and 33 Seyfert 2s ($\sim$20\% of the total number with 
spectroscopic information in the original sample). These include some of the most extreme warm sources.
In order to establish the true relations between IR and optical properties we need 
to compare our results for the Warm sample with those for a control sample of 
galaxies selected to have ``cold'' mid-IR colours, \ie the host galaxy dominating 
over the nucleus. Hence, we carried out similar observations for a smaller sample 
(16 objects) of Cold IRAS sources, selected to have $\alpha_{(25,60)}\leq$-1.5 
and to span the same redshift and $L_{IR}$ range as the Warm sample (Figure~\ref{f2}). Fourteen of the objects in our Warm subsample are Seyfert 2 galaxies
with $L_{FIR}\geq 10^{11}\Lsun$ ($\sim$50\% of the total number in the original
sample) and four of them have $L_{60}\geq 10^{11}\Lsun$ (and 
$L_{8-1000 \mu m}\geq 10^{12}\Lsun$) (indicated with an asterisk in our tables).
Investigation of these powerful galaxies is relevant for determining the origin of 
the ``QSO 2 deficit'' (paucity of high-luminosity narrow-line AGNs), as they lie 
close to the presumed maximum limiting luminosity for narrow-line AGNs. 

In Tables~\ref{tab1} and~\ref{tab2} we give some basic information for the objects in
our Warm and Cold samples, respectively, listed in order of (decreasing) 
mid-IR (25-60 $\mu$m) warmness. For each object we list: the IRAS name, spectral 
type, redshift, far-IR luminosity and mid- and far-IR spectral indices. Together 
with the IRAS name we give the serial number from the original sample 
(\cite{degrijp2}). Although for some of these objects newr spectroscopic data
were available, for uniformity we use the spectral types and redshifts from 
\cite{degrijp3} for all our objects.
An important point here, concerns the ``real'' Seyfert type of our objects.
In some cases (\eg IRAS 11365-3727 (I286) and 13536+1836 (I333)) it was shown
that optically classified Seyfert 2 nuclei show Seyfert 1-like broad line 
spectra in polarized light. For these known cases we chose to use their optical
(that is, Seyfert 2) identification for reasons of uniformity, as it is likely 
that there are more (not known) cases of obscured Seyfert 1 nuclei in our 
sample. Testing the obscuration effect, using the IR properties of 
optically-classified Seyferts, is one of the aims of this study and having a 
few confirmed cases might actually help to interpret our results.

\placetable{tab1}

\placetable{tab2}

\begin{figure}
\epsscale{.8}
\plotone{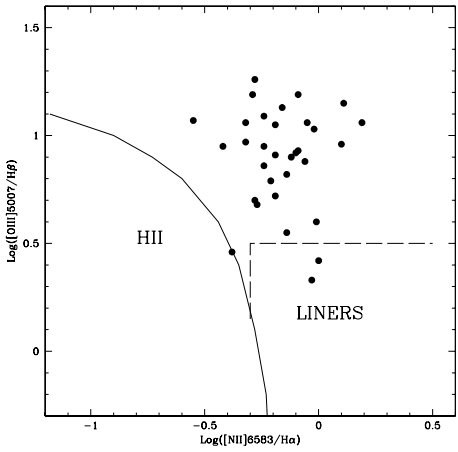}
\caption{One of the diagnostic diagrams of Veilleux and Osterbrock 1987 for the Warm Seyfert 2 galaxies in the present sample. The full line indicates the boundaries between \ion{H}{2} and Seyfert loci and the broken line the boundaries between Seyfert and LINER loci. \label{f1}}
   \end{figure}

Another source of concern is the possible partial overlap in optical spectral 
properties between Seyferts, LINERS and starbursts that might affect the Seyfert 
Warm sample selection. \cite{degrijp3} classified the 
emission-line spectra using the spectral classification criteria introduced by 
\cite{veilleux87}. The latter established a classification system for the emission 
line galaxies, that involves the emission line ratios [\ion{O}{3}]$_{5007}/{H_{\beta}}$,
[\ion{N}{2}]$_{6583}/H_{\alpha}$, [\ion{S}{2}]$_{6716+6731}/H_{\alpha}$
and [\ion{O}{1}]$_{6300}/H_{\alpha}$ and was applied by \cite{degrijp3}
 in order to classify the (nuclear) spectra of their Warm sample. In
fact, all nuclei with broader permitted than forbidden lines were classified 
as Seyfert type 1, while nuclei showing ratios [\ion{O}{3}]$_{5007}/H_{\beta} >$3 
and [\ion{N}{2}]$_{6583}/H_{\alpha} >$0.5 were classified as 
Seyfert type 2. Spectra with ratios [\ion{O}{3}]$_{5007}/H_{\beta} <$3 
and [\ion{N}{2}]$_{6583}/H_{\alpha} >$0.5 were ascribed to the LINER category and the
remaining objects  with line ratios outside the above ranges, were classified as 
\ion{H}{2} or starburst type (\cite{degrijp3}). It is clear that this classification is an 
over-simplification. As Veilleux \& Osterbrock (\cite{veilleux87}) have discussed 
extensively, determining such boundaries is not without uncertainties, especially if
only two line ratios are used at a time and the nature of objects in any 
``transition zone'' is highly uncertain. These deficiencies were acknowledged by 
\cite{degrijp3} who in a few cases have also employed the ratio 
[\ion{O}{1}]$_{6300}/H_{\alpha}$, to clarify doubtful classifications. 
A misclassification of the order of $\sim$5\% (Terlevich R. 1995, private 
communication) is estimated between Seyfert 2, LINERs and starbursts. For this 
reason, we have checked the nuclear spectral type of our objects, whenever another 
source of classification appears in the literature. In Figure~\ref{f1} we 
have reconstructed one of the diagnostic diagrams of \cite{veilleux87} for the 
Seyfert 2 galaxies in the present sample. We indicate the boundaries between Seyfer
and \ion{H}{2}-region like objects with a full line and between Seyfert and LINERs with a 
broken line. Most of our galaxies indeed fall within the AGN area but
one object, IRAS 01346-0924 (I28), has a possible \ion{H}{2}-like spectrum
and two, IRAS 03278-4329 (I90) and IRAS 19254-7245 (I489) have possible LINER
spectra. The Seyfert 1 galaxies are not plotted here, because the line ratios
published in \cite{degrijp3} include both the broad and narrow-line components.
Moreover, the De Grijp \etal Seyfert 1 classification, as described above, is a 
broad-line-region type classification, which encompasses various AGN classes with a 
range of properties: Seyfert 1 to Seyfert 1.9 types and maybe broad line radio 
galaxies (BLRGs). To distinguish between these classes, one needs to observe other 
emission lines (\eg \ion{Fe}{2}$_{4570,5250}$) and to know their radio properties.

\placefigure{f1}

The optical spectral classification of \cite{degrijp3} might have introduced a 
selection effect towards more powerful Seyfert 2 galaxies. This is because, while 
the selection of Seyfert 1s was based on the detection of a broad emission line 
component -independently of ionization strength-, Seyfert 2s were selected only if 
their line ratios were indicating gas predominantly photoionized by the AGN. This 
means that, within the orientation/obscuration scenario, Seyfert 2s with less 
powerful AGNs, smaller NLRs and possibly strong star formation, might have been 
misclassified (as starbursts or LINERs) if their AGN line emission was masked by the
lower ionization (\ion{H}{2}- or LINER-like)
line ratios. The same objects however, would have been selected (as Seyfert 1s)
if they were viewed face-on. This possible bias is only an issue {\em if} the 
orientation/obscuration scenario is correct. However, there are two reasons
why we believe that bias should be minimal if existent at all, in the present 
sample: (i) As we pointed out in Section 1, the IR flux selection criterion 
would pick up Seyferts with high bolometric luminosities, their energy
output peaking in the infrared. Thus, it is unlikely that many weak AGNs are
included in the present Seyfert sample. (ii) The lower limit in the mid-IR colour 
criterion tends to eliminate the low luminosity AGNs, because the smaller 
nuclear/host galaxy relative dominance tends to cool down their mid-IR colours. This
exclusion will affect similarly Seyfert 1s and 2s, unless the dusty torus is 
optically thick even at 25 $\mu$m. However this does not seem to be the case, at 
least for this sample: \cite{degrijp3} have shown that the difference in
median spectral shape between Seyfert 1 and 2 nuclei in their ``warm'' sample is very small:
$\lesssim$0.12 between 25 and 60 $\mu$m and $\gtrsim$0.10 between 60 and 100 $\mu$m.

The far-IR luminosity $L_{FIR}$ listed in Tables~\ref{tab1} and~\ref{tab2} is 
calculated from the formula 
\[ FIR = 1.26\times 10^{-11} (2.58 f_{60} + f_{100}) R \]  (\cite{helou88}) for H$_{o}$=75 km
sec$^{-1}$ Mpc$^{-1}$ and q$_{o}$=0. {\em FIR} is the 60-100 $\mu$m far-IR flux 
density in ergs cm$^{-2}$ sec$^{-1}$ and $f_{60},f_{100}$ are the 
monochromatic IRAS flux densities in Jy. The correction factor R accounts for the
flux missed longward of the IRAS 100 $\mu$m band and is taken from \cite{lonsdale85},
assuming a power law $\sim\nu^{1}$ for the dust emissivity.\\ 

The IR colour indices are calculated adopting a flux density
$\propto\nu^{\alpha}$, from the relations:
\[\alpha_{(25,60)} = 2.63 \log{\frac{f_{25}}{f_{60}}}\]
and
\[\alpha_{(60,100)} = 4.51 \log{\frac{f_{60}}{f_{100}}}\]

The flux densities used to calculate IR luminosities and colour indices
are from the co-added IRAS data (\cite{degrijp3}).

\placefigure{f2}

In Figure~\ref{f2} we show the distributions of: redshift, log(L$_{FIR}$), 
$\alpha_{(25,60)}$ and $\alpha_{(60,100)}$, for the Warm Seyfert sample, split in two 
subsamples for the type 1 and 2 Seyferts and for the Cold sample.
The calculated median values for each (sub)sample are indicated as vertical bars on 
the lower x-axes and are listed in Table~\ref{tab3} together with the means and 
standard deviations. 
The redshift distribution is similar for the Seyfert 1 and Cold samples,
in the range 0.01-0.08. The F-test and Student's t-test show no significantly 
different variances and means for the two distributions. Seyfert 2s have a similar 
distribution but is shifted to higher z values, mainly because of the presence of 
the high-$L_{FIR}$ objects in the sample. The three samples span similar ranges in 
far-IR luminosities, with a slight tendency of the Seyfert 1 sample towards lower 
$L_{FIR}$ and of the Cold sample towards higher $L_{FIR}$. A Student's t-test shows 
that the difference between the means of the Warm Seyfert 1 and 2 samples and 
between the means of the Seyfert 1 and Cold samples is statistically significant
(at the 0.05 and 0.008 significance level, respectively). On the other hand, the 
Seyfert 2 and Cold samples do not have significantly different variances or means. 
A K-S test shows that
none of the three samples match one another at a statistically significant level.
The tendency of Seyfert 2s in this sample to show higher IR luminosities could
be attributed to the possible bias towards more powerful AGNs in these objects, as
discussed earlier. However, we will see later in a forthcoming paper that this is 
rather related to the larger fraction of interacting systems, among the Seyfert 2
galaxies. 

\begin{figure}
\epsscale{1.2}
\plottwo{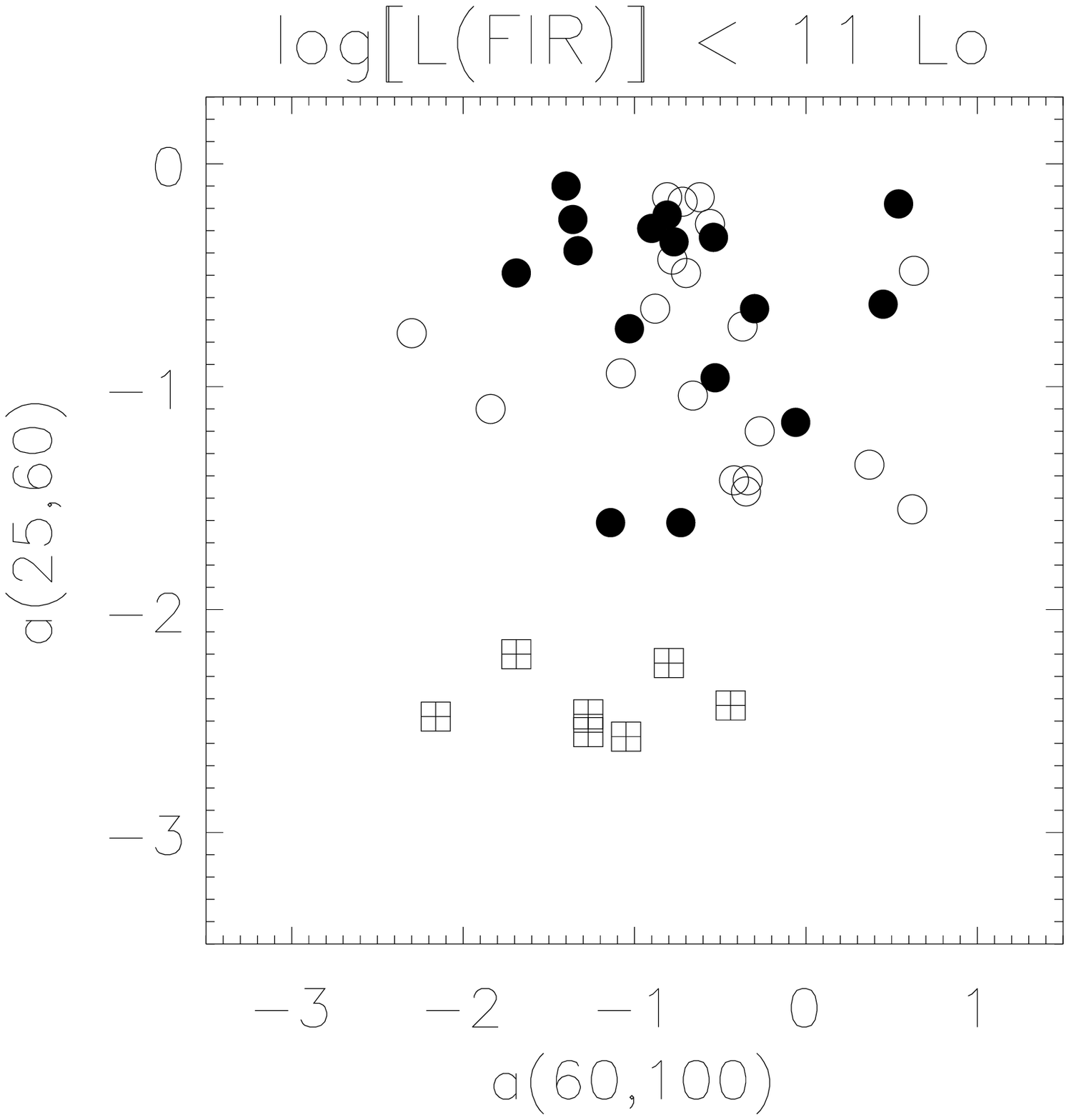}{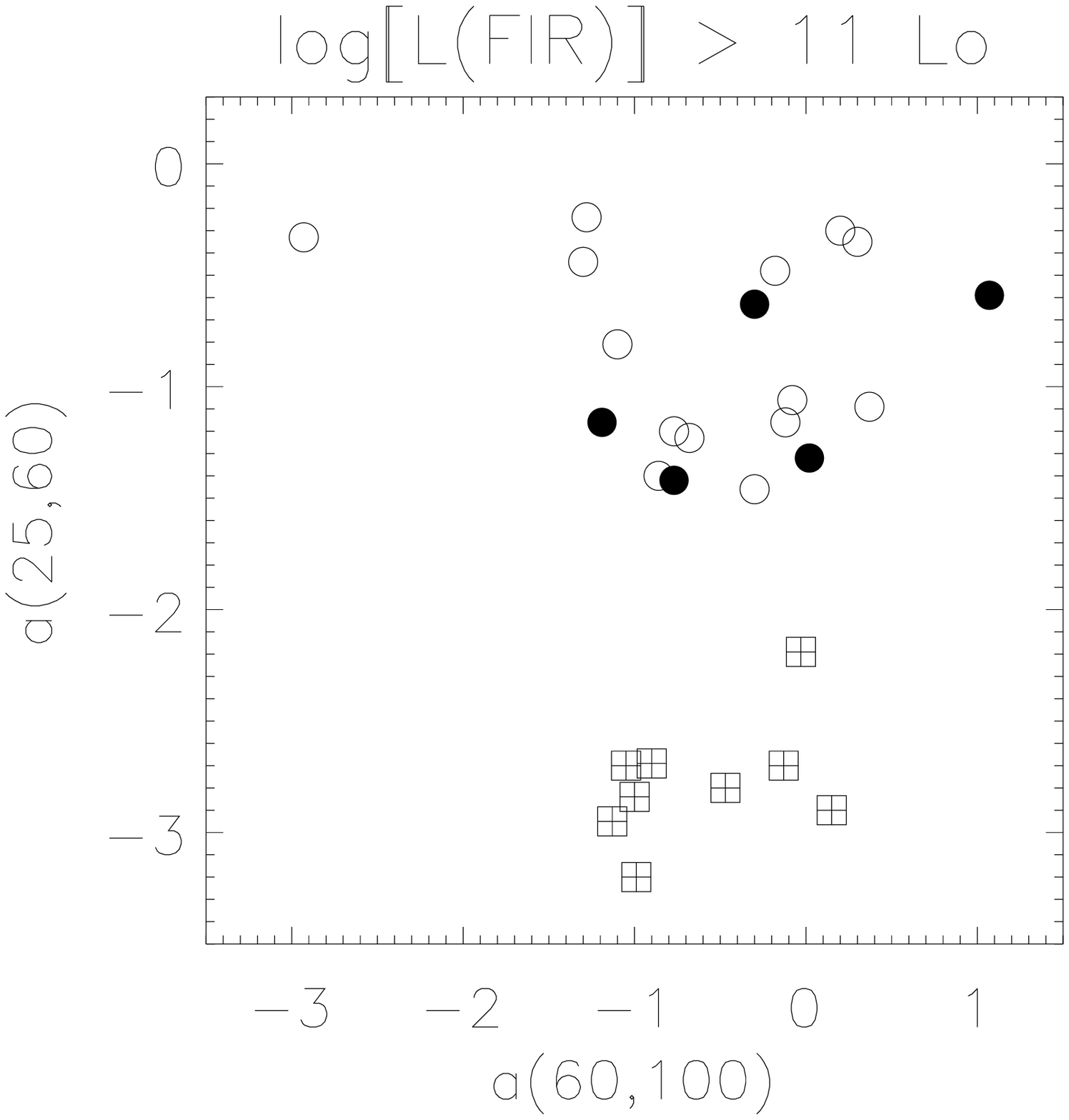}
\caption{IR colour-colour diagrams for two bins of far-IR (60-100 $\mu$m) luminosities. Filled circles represent Warm Seyfert 1s, open circles Warm Seyfert 2s and crossed squares Cold galaxies. \label{f3}}
   \end{figure}

In Figure~\ref{f3} we show the IR colour-colour plots in 
two bins of L$_{FIR}$, using different symbols for the three (sub)samples.
The 25-60 $\mu$m colour index was the main selection criterion for our Warm
and Cold samples. The two Warm Seyfert samples span the same range of 
$\alpha_{(25,60)}$ and there is no statistically significant difference in their 
variances and means. 
The 60/100 flux ratio is an indicator of the dominance of the warm dust 
component, due mainly to star formation. The Warm and Cold 
samples span a similar range in $\alpha_{(60,100)}$, with the latter shifted to
somewhat colder far-IR colours. In fact, a K-S test shows that the Cold and Seyfert 2
samples have significantly different distributions (the null hypothesis that they are similar can be rejected at the 95\% significance level). We will analyse these further in correlation with our photometry results, in Paper II.

\placefigure{f3}

In Table~\ref{tab3} we include for comparison mean values for additional 
IR-selected samples: (a) a sample of isolated galaxies, drawn from the complete IRAS
bright galaxy sample (with $f_{60}\geq$5.24 Jy; \cite{soifer2}) (b) a sample of 
powerful FIR galaxies selected to have warm 60-100 $\mu$m colours (\cite{armus90}) and 
(c) the ULFIRG 
sample (L$_{FIR}\geq$10$^{12}$\Lsun; \cite{sanders1}). The latter two samples contain
strongly interacting/merging systems, undergoing strong star formation events. 
Our Warm Seyferts do not match any of the above three samples: they have larger IR 
luminosities and warmer mid- and far-IR colours than the isolated IR-bright 
galaxies, while fainter luminosities and colder far-IR colours but warmer mid-IR 
colours than the two powerful-IR galaxy samples. Our Cold control sample is more 
comparable to the Armus \etal ``cold'' sample, but with redder far-IR colours.
From this quick comparison we conclude that although the above samples contain objects with
similar or higher IR luminosities as our objects, their mid-IR colours are 
generally colder than our sample objects.  

\placetable{tab3}

\section{Observations}

The observations were carried out over a period of several years mostly with the 
telescopes of ESO at La Silla and for a few objects with the CFHT at Mauna Kea. The 
ESO observations were carried out with the Danish 1.54m and Dutch 0.9m telescopes.
Warm Seyferts were observed in order of 25-60 $\mu$m warmness in the De Grijp \etal 
sample and Cold objects in order of 25-60 $\mu$m coldness in the IRAS Catalogue of 
Galaxies (\cite{lonsdale85}). Here we present data for those objects for which we 
have the best deep images. The observations consist of imaging in several or all of 
the $B,V,R,I$ bands and, for
about a third of all objects, also in the $H\alpha$ and/or [\ion{O}{3}] narrow bands.
On La Silla we have used the Bessel $B$ (\#419, \#450), $V$ (\#420, \#451),
$R$ (\#421, \#452) and Gunn $i$ (\#465, \#425) filters and at the CFHT $B$ 
(\#4402), $V$ (\#4504), $R$ (\#4609) filters that (as the La Silla filters) correspond to 
the Johnson photometric system. Consequently, our measured magnitudes from different
telescopes do not need conversion between photometric systems.
The log of the observations, instrumental configurations, exposure times and seeing 
conditions are given in Tables~\ref{tab4} and~\ref{tab5}.

The field size and spatial sampling vary among objects, because the instrumentation 
and the CCDs changed over time, even for the same telescope. This added to the 
complication of producing a consistent photometric calibration, as one cannot 
directly intercompare calibration curves from different observing sessions.
 Our photometric calibrations are 
based on observations of several standard stars per session. However, for a few 
sessions with bad weather conditions, no calibration was applied. Standard reduction procedures were used to apply the bias and dark subtractions and 
for the flat-field and cosmic-ray corrections.

In the final published version of this paper we will present broad-band 
photometric contour maps for all our observed (54 Warm and 16 Cold) objects
(omitted in this preprint version, to reduce the file size). Within each (sub)sample the objects are
presented in order of increasing IRAS number: Seyfert 1: IRAS 00509+1225,
IRAS 23016+2221 Seyfert 2: IRAS 00198-7926, IRAS 23254+0830, Cold: IRAS 02439-7455,
IRAS 23179-6929. The filter ID is indicated on the lower right corner of each 
figure. Due to the fact that the images were taken with various telescopes and CCDs,
the spatial sampling varies between figures. The scale is indicated by crosses of 
5\arcsec on the upped right corner of the contour maps. The centers (0,0) indicate 
the target galaxy position, except in cases of closely interacting or merging 
systems, where the IRAS source cannot be uniquely identified (see also Paper II).
All brightness contours were plotted at the levels 2$\sigma$-10$^{3}\sigma$
as: 2$\sigma$, 20$\sigma$, 30$\sigma$, 50$\sigma$, 100$\sigma$, 200$\sigma$, 
300$\sigma$, 500$\sigma$, (1000$\sigma$). Exceptions to this are two shallow 
images: IRAS 03278-4329 (25$\sigma$) and IRAS 05207-2727 (5$\sigma$).

\section{Concluding Remarks}

In the present paper (Paper I) we have presented an overview of the data that 
will be analyzed and presented in detail in four forthcoming papers.
In these papers we shall parametrize and discuss the luminosities, colours, gradients, sizes and 
morphologies for our sample objects. These optical properties will be compared to IR properties 
and the three (sub)samples will be inter-compared. 
We shall concentrate on two questions: (i) Based on an isotropically IR-selected 
sample, can we test the orientation unification scheme for Seyferts? (ii) Is the 
origin of the mid-IR excess in Warm Seyferts thermal or non-thermal? AGN or 
starburst-dominated? In particular, we will investigate (a) the distribution of IR 
(isotropic) properties and how they correlate with optical (anisotropic) properties 
(b) the host galaxy effects (morphologies and dimensions) (c) the connection between
starburst, AGN and IR activity (d) the dust extinction effects and (e) the 
connection between interaction stage and nuclear/IR activity. In Paper II we will 
present and discuss the results of aperture photometry. Papers III and IV will be 
dedicated to surface photometry in terms of light and colour profiles, respectively. 
Finally, in 
Paper V we will discuss all the results in the context of galactic interactions.

\placetable{tab4}

\placetable{tab5}

\acknowledgments
I am grateful to my thesis advisors George Miley and Walter Jaffe for providing
me with stimulation and support throughout the completion of this project. I 
am also grateful to J. Gerritsen, van Hemert, R. de Grijs, H. Verkouter and P. 
Groot for their help in collecting some of the data presented here.
This research has made use of the NASA/IPAC Extragalactic Database (NED) 
which is operated by the Jet Propulsion Laboratory, California Institute of 
Technology, under contract with the National Aeronautics and Space 
Administration. Part of this work was completed while the author held a 
National Research Council - NASA GSFC Research Associateship.

%

%
%

%
%

\clearpage

\begin{table}
\dummytable\label{tab1}
\end{table}

\clearpage

\begin{deluxetable}{lrrrrr}
\tablecaption{IR Properties: Cold Sources. \label{tab2}}
\tablewidth{0pt}
\tablehead{
\colhead{Identification} & \colhead{Sp. Type} & \colhead{Redshift} & 
\colhead{log(L$_{FIR}$)\tablenotemark{1}} & \colhead{$\alpha_{(25,60)}$} & \colhead{$\alpha_{(60,100)}$} \\
\colhead{} & \colhead{} & \colhead{} & \colhead{$\Lsun$} & \colhead{} & \colhead{} 
}
\startdata
            IRAS 23128-5919 (I731) & S2\tablenotemark{2} & 0.0446\tablenotemark{3} & 11.88 & -2.19 & -0.03 \nl
            IRAS 07514+5327 (I231) & \ion{H}{2}\tablenotemark{2} & 0.0248\tablenotemark{2} & 10.43 & -2.20 & -1.69 \nl
            IRAS 06506+5025 (I211) & S1\tablenotemark{2} & 0.0200\tablenotemark{2} & 10.27 & -2.24 & -0.80 \nl 
            IRAS 02439-7455 & normal\tablenotemark{4} & 0.0329\tablenotemark{5} & 10.50 & -2.43 & -0.44 \nl
            IRAS 04015-1118 & \nodata & 0.0296\tablenotemark{6} & 10.98 & -2.47 & -1.27 \nl
            IRAS 05217-4245 & (\ion{H}{2})\tablenotemark{7} & 0.0343\tablenotemark{7} & 10.95 & -2.48 & -2.16 \nl
            IRAS 04265-4801 & S2 & 0.0165\tablenotemark{3}  & 10.31 & -2.55 & -1.27 \nl
            IRAS 10475-1429 & \nodata & 0.0259\tablenotemark{8} & 10.78 & -2.57 & -1.05 \nl
            IRAS 09406+1018 & \nodata & 0.0538\tablenotemark{9} & 11.43 & -2.69 & -0.90 \nl
            IRAS 04530-3850 & \ion{H}{2}\tablenotemark{7} & 0.0453\tablenotemark{7} & 11.51 & -2.70 & -1.05 \nl
            IRAS 19184-7404 (I712) & S2\tablenotemark{2} & 0.0702\tablenotemark{10} & 11.48 & -2.70 & -0.13  \nl
            IRAS 03531-4507 & \nodata & 0.0514\tablenotemark{9} & 11.26 & -2.80 & -0.47 \nl
            IRAS 04304-5323 & \nodata & 0.0588\tablenotemark{11} & 11.25 & -2.84 & -1.00 \nl
            IRAS 04454-4838 & \nodata & 0.0529\tablenotemark{12} & 11.75 & -2.90 &  0.15 \nl
            IRAS 23179-6929 & (\ion{H}{2}) & 0.0416\tablenotemark{8}  & 11.59 & -2.95 & -1.13 \nl
            IRAS 05207-2727 & \nodata & 0.0341\tablenotemark{8}  & 11.04 & -3.20 & -0.99 \nl
\enddata
\tablerefs{
$^{1}$ \cite{FSC}, $^{2}$ \cite{degrijp3}, $^{3}$ \cite{lauberts}, $^{4}$ \cite{fairall1}, $^{5}$ \cite{3dRC}, $^{6}$ \cite{huchra}, $^{7}$ \cite{segi}, $^{8}$ \cite{straus}, $^{9}$ \cite{cfa}, $^{10}$ \cite{fairall2}, $^{11}$ \cite{fisher}, $^{12}$ \cite{sanders2}}
\tablecomments{Uncertainties for the IRAS fluxes can be found in \cite{lonsdale85}}
\end{deluxetable}

\clearpage

\begin{deluxetable}{rrrrr}
\tablecaption{Median and Mean Values and Standard Deviations. \label{tab3}}
\tablehead{
\colhead{} & \colhead{$z$} & \colhead{L$_{FIR}$} & \colhead{$\alpha_{(25,60)}$} & \colhead{$\alpha_{(60,100)}$}
}
\startdata
\cutinhead{Warm Seyfert 1}
Median  & 0.0385 & 10.70 & -0.63 & -0.77 \nl
Mean    & 0.0432 & 10.86 & -0.72 & -0.61 \nl
$\sigma$ & 0.020 & 0.44 & 0.48 & 0.71 \nl
\cutinhead{Warm Seyfert 2}
Median  &  0.0485 & 10.93 & -0.81 & -0.62 \nl
Mean    &  0.0520 & 11.16 & -0.83 & -0.60 \nl
$\sigma$ & 0.024 & 0.47 & 0.47 & 0.77 \nl     
\cutinhead{Cold Sample}
Median  & 0.0416 & 11.25 & -2.57 & -0.99 \nl
Mean    & 0.0404 & 11.33 & -2.62 & -0.89 \nl
$\sigma$ & 0.014 & 0.49 & 0.28 & 0.60 \nl
\cutinhead{Comparison Samples - Mean values}
IR Isol\tablenotemark{1} & & 10.30 & -2.37 & -1.49 \\                    
IR CS\tablenotemark{2} & & 11.36 & $\leq$-1.5 & $\geq$-0.5 \nl           
ULIRGs\tablenotemark{3} & & 11.90 & -2.35 & 0.09 \nl                        
\enddata
\tablerefs{
$^{1}$ IRAS bright isolated galaxies \cite{soifer2},
$^{2}$ Powerful IR Cold galaxies \cite{armus90},
$^{3}$ Ultra-luminous IR galaxies \cite{sanders1}}
\end{deluxetable}

\clearpage

\begin{table}
\dummytable\label{tab4}
\end{table}

\begin{table}
\dummytable\label{tab5}
\end{table}

%
%

\end{document}


\begin{deluxetable}{lrrrrrr}
\tablenum{4}
\tablecaption{Observing Log: Warm Sources. \label{tab4}}
\tablewidth{0pt}
\tablehead{
\colhead{Identification} & \colhead{Date} & \colhead{Tel-Instr-CCD} & 
\colhead{Spatial sampl.} & \colhead{Filter} & 
\colhead{Exp.Time\tablenotemark{1}} & \colhead{Seeing\tablenotemark{2}} \\
\colhead{} & \colhead{} &  \colhead{} & \colhead{(arcsec/pix)} & \colhead{} &
\colhead{(sec)} & \colhead{(arcsec)}
}
\startdata
            IRAS 15015+1037 (I359) & 04/93 & 0.9m Dutch-Adapter-No33 & 0.443 & $B$ & 1200 & 1.5 \nl
                                   &       &                  &       & $V$ & 600  & 1.7 \nl
                                   &       &                  &       & $R$ & 600  & 1.4 \nl
                                   &       &                  &       & $I$ & 600  & 1.4 \nl
            IRAS 04507+0358 (I156) & 02/92 & 0.9m Dutch-Adapter-No7 & 0.36 & $B$ & 420 & 1.7 \nl
                                   &       &                 &      & $V$ & 240 & 1.5 \nl
                                   &       &                 &      & $R$ & 240 & 1.3 \nl
                                  &       &                 &      & $I$ & 240 & 1.4 \nl
                                   & 12/95 & 1.54m Danish-DFOSC-Loral & 0.39 & $B$ & 1200 & 1.6 \nl
                                   &       &                      &      & $R$ & 240  & 1.4 \nl
            IRAS 02580-1136 (I67)  & 10/95 & 0.9m Dutch-Adapter-No33 & 0.443 & $V$ & 360 & 1.5 \nl
                                   &       &                  &       & $R$ & 300 & 1.4 \nl
                                   & 12/95 & 1.54m Danish-DFOSC-Loral & 0.39 & $B$ & 600 & 1.4 \nl
                                   &       &                      &      & $R$ & 600 & 1.2 \nl 
            IRAS 03202-5150 (I80)  & 08/92 & 0.9m Dutch-Adapter-No7 & 0.36 & $B$ & 2700 & 1.5 \nl
                                   &       &                 &      & $V$ & 900  & 1.5 \nl
                                   &       &                 &      & $R$ & 900  & 1.5 \nl
                                   &       &                 &      & $I$ & 900  & 1.4 \nl
                                   & 02/92 &                 &      & $H\alpha$ & 1800 & 1.4 \nl
                                   & 10/95 & 0.9m Dutch-Adapter-No33 & 0.443 & $V$ & 600 & 1.2 \nl
                                   &       &                  &       & $R$ & 420 & 1.0 \nl
                                   &       &                  &       & Ha & 1800 & 1.1 \nl
            IRAS 09497-0122 (I260) & 03/92 & 3.6m CFHT-SILFID-Lick2 & 0.178 & $V$ & 180 & 1.1 \nl
                                   &       &                        &       & $R$ & 210 & 0.8 \nl
            IRAS 04124-0803 (I114) & 02/92 & 0.9m Dutch-Adapter-No7 & 0.36 & $B$ & 420 & 1.8 \nl
                                   &       &                 &      & $V$ & 240 & 1.6 \nl
                                   &       &                 &      & $R$ & 240 & 1.5 \nl
                                   &       &                 &      & $I$ & 240 & 1.5 \nl
                                   &       &                 &      & $H\alpha$ & 1800 & 1.7 \nl
                                   & 12/95 & 1.54 Danish-DFOSC-Loral & 0.39 & $B$ & 600 & 1.4 \nl
                                   &       &                     &      & $R$ & 90  & 1.3 \nl
            IRAS 09435-1307 (I256) & 04/95 & 1.54 Danish-DFOSC-No17 & 0.51 & $I$ & 600 & 1.2 \nl
            IRAS 13512-3731 (I330) & 02/92 & 0.9 Dutch-Adapter-No7 & 0.36 & $B$ & 420 & 1.3 \nl
                                   &       &                &      & $V$ & 240 & 1.3 \nl
                                   &       &                &      & $R$ & 240 & 1.1 \nl
                                   &       &                &      & $I$ & 240 & 1.0 \nl
            IRAS 15304+3017 (I375) & 07/93 & 0.9 Dutch-Adapter-No33 & 0.443 & $V$ & 600 & 1.8 \nl
                                   &       &                 &       & $R$ & 600 & 1.9 \nl
                                   &       &                 &       & $I$ & 600 & 2.2 \nl
            IRAS 04493-6441 (I153) & 10/95 & 0.9 Dutch-Adapter-No33 & 0.443 & $V$ & 1200 & 1.6 \nl
                                   &       &                 &       & $R$ & 900 & 1.4 \nl
                                   & 12/95 & 1.54 Danish-DFOSC-Loral & 0.39 & $B$ & 1800 & 1.4 \nl
                                   &       &                     &      & $R$ & 480  & 1.3 \nl
            IRAS 00521-7054 (I19)  & 12/95 & 1.54 Danish-DFOSC-Loral & 0.39 & $B$ & 1800 & 1.4 \nl
                                   &       &                     &      & $R$ & 450  & 1.3 \nl
                                   & 08/92 & 0.9m Dutch-Adapter-No7 & 0.36 & $V$ & 600 & 1.8 \nl
                                   &       &                        &      & $R$ & 600 & 1.9 \nl
                                   &       &                        &      & $I$ & 600 & 2.2 \nl
            IRAS 09305-8408 (I254) & 04/93 & 0.9m Dutch-Adapter-No33 & 0.443 & $B$ & 900 & 1.9 \nl
                                   &       &                  &       & $V$ & 600 & 1.7 \nl
                                   &       &                  &       & $R$ & 600 & 2.0 \nl
                                   &       &                  &       & $I$ & 600 & 2.1 \nl
                                   & 04/95 & 1.54 Danish-DFOSC-No17 & 0.51 & $I$ & 60 & 1.1\nl
            IRAS 19580-1818 (I495) & 05/92 & 0.9 Dutch-Adapter-No7 & 0.36 & $B$ & 600 & 1.5 \nl
                                   &       &                &      & $V$ & 240 & 1.2 \nl
                                   &       &                &      & $R$ & 240 & 1.3 \nl
                                   &       &                &      & $I$ & 240 & 1.2 \nl
                                   &       &                &      & $H\alpha$ & 1800 & 1.3 \nl
                                   &       &                &      & [\ion{O}{3}] & 1800 & 1.2 \nl
                                   & 04/95 & 1.54 Danish-DFOSC-No17 & 0.51 & $V$ & 300 & 1.1 \nl
            IRAS 13536+1836 (I333) & 03/92 & 3.6m CFHT-SILFID-Lick2 & 0.178 & $B$ & 480 & 1.0 \nl
                                   &       &                  &       & $V$ & 240 & 0.8 \nl
                                   &       &                  &       & $R$ & 240 & 0.7 \nl
                                   &       &                  &       & $H\alpha$\tablenotemark{3} & 1800 & 1.4\tablenotemark{4} \nl
                                   &       &                  &       & [\ion{O}{3}]\tablenotemark{3} & 1800 & 1.4\tablenotemark{4} \nl
            IRAS 11365-3727 (I286) & 04/95 & 1.54 Danish-DFOSC-No17 & 0.51 & $I$ & 30 & 1.1 \nl
            IRAS 05136-0012 (I171) & 10/95 & 0.9m Dutch-Adapter-No33 & 0.443 & $V$ & 150 & 1.5 \nl
                                   &       &                  &       & $R$ & 60  & 1.4 \nl
                                   &       &                        &      & $I$ & 600 & 2.2 \nl
            IRAS 03355+0104 (I96)  & 12/95 & 1.54 Danish-DFOSC-Loral & 0.39 & $B$ & 180 & 1.6 \nl
                                   &       &                     &      & $R$ & 30  & 1.2 \nl
            IRAS 13144+4508 (I315) & 03/92 & 3.6m CFHT-SILFID-Lick2 & 0.178 & $B$ & 600 & 1.1\nl
                                   &       &                         &      & $V$ & 300 & 0.8 \nl
                                   &       &                         &      & $R$ & 300 & 1.0 \nl
                                   &       &                  &       & $H\alpha$\tablenotemark{3} & 1800 & 0.7\tablenotemark{4} \nl
            IRAS 22017+0319 (I528) & 10/95 & 0.9m Dutch-Adapter-No33 & 0.443 & $V$ & 1200 & 2.0 \nl
                                   &       &                        &       & $R$ & 750  & 1.6 \nl
            IRAS 20481-5715 (I512) & 08/92 & 1.54m Danish-Adapter-No17 & 0.298 & $B$ & 420 & 1.4 \nl
                                   &       &                         &       & $V$ & 240 & 1.3 \nl
                                   &       &                         &       & $R$ & 240 & 1.3 \nl
                                   &       &                         &       & $I$ & 240 & 1.4 \nl
                                   &       &                         &       & $H\alpha$ & 1800 & 1.3 \nl
            IRAS 05218-1212 (I176) & 04/95 & 1.54m Danish-DFOSC-No17 & 0.51 & $V$ & 120 & 1.3 \nl
                                   &       &                        &       & $I$ & 180 & 1.5 \nl
            IRAS 09182-0750 (I253) & 04/95 & 1.54m Danish-DFOSC-No17 & 0.51 & $V$ & 420 & 1.2 \nl
                                   &       &                        &       & $I$ & 600 & 1.1 \nl
            IRAS 01378-2230 (I31)  & 09/95 & 0.9m Dutch-Adapter-No33 & 0.443 & $B$ & 1350 & 2.3 \nl
                                   &       &                        &       & $R$ & 510  & 1.9 \nl
            IRAS 00509+1225 (I18)  & 08/92 & 1.54m Danish-Adapter-No17 & 0.298 & $B$ & 180 & 1.7 \nl
                                   &       &                          &       & $V$ & 60  & 1.4 \nl
                                   &       &                          &       & $R$ & 60  & 1.4 \nl
                                   &       &                          &       & $I$ & 60  & 1.7 \nl
            IRAS 21299+0954 (I521) & 08/92 & 1.54m Danish-Adapter-No17 & 0.298 & $B$ & 428 & 1.3 \nl
                                   &       &                          &       & $V$ & 120 & 1.3 \nl
                                   &       &                          &       & $R$ & 60 & 1.9 \nl
                                   &       &                          &       & $I$ & 30 & 1.5 \nl
                                   &       &                          &       & $H\alpha$ & 1200 & 1.7 \nl
                                   &       &                          &       & [\ion{O}{3}] & 1200 & 1.8 \nl
            IRAS 14557-2830 (I358) & 08/95 & 1.54m Danish-DFOSC-No17 & 0.51 & $V$ & 450 & 2.3 \nl
                                   &       &                        &      & $I$ & 450 & 2.0 \nl
            IRAS 11249-2859 (I282) & 04/95 & 1.54m Danish-DFOSC-No17 & 0.51 & $I$ & 300 & 1.2 \nl
            IRAS 03059-2309 (I72)  & 08/92 & 1.54m Danish-Adapter-No17 & 0.298 & $B$ & 900 & 2.1 \nl
                                   &       &                          &       & $V$ & 420 & 1.2 \nl
                                   &       &                          &       & $R$ & 600 & 1.3 \nl
                                   &       &                          &       & $I$ & 300 & 1.4 \nl
                                   &       &                          &       & $H\alpha$ & 900 & 1.2 \nl
                                   &       &                          &       & [\ion{O}{3}] & 900 & 1.3 \nl
            IRAS 11215-2806 (I281) & 04/95 & 1.54m Danish-DFOSC-No17 & 0.51 & $I$ & 300 & 1.2 \nl
            IRAS 03362-1641 (I98)  & 12/95 & 1.54m Danish-DFOSC-Loral & 0.39 & $B$ & 600 & 2.1 \nl
                                   &       &                          &      & $R$ & 340 & 1.8 \nl
            IRAS 15599+0206 (I392) & 04/95 & 1.54m Danish-DFOSC-No17 & 0.51 & $V$ & 120 & 1.2 \nl
                                   &       &                        &      & $I$ & 180 & 1.2 \nl
            IRAS 06317-6403 (I206) & 04/95 & 1.54m Danish-DFOSC-No17 & 0.51 & $V$ & 420 & 1.5 \nl
            IRAS 18401-6225 (I473) & 04/95 & 1.54m Danish-DFOSC-No17 & 0.51 & $V$ & 150 & 1.5 \nl
            IRAS 00598-1956 (I20)  & 09/95 & 0.9m Dutch-Adapter-No33 & 0.443 & $B$ & 1350 & 2.0 \nl
                                   &       &                        &       & $R$ & 800  & 2.0 \nl
                                   &       &                        &       & $H\alpha$ & 1800 & 1.9 \nl
            IRAS 03335-5625 (I94)  & 08/95 & 1.54m Danish-DFOSC-No17 & 0.51 & $V$ & 600 & 2.5 \nl
                                   &       &                        &      & $R$ & 300 & 2.0 \nl
                                   &       &                        &      & $I$ & 600 & 2.4 \nl
                                   &       &                        &      & $H\alpha$ & 1800 & 2.2 \nl
            IRAS 00198-7926 (I6)   & 08/95 & 1.54m Danish-DFOSC-No17 & 0.51 & $V$ & 600 & 2.7 \nl
                                   &       &                        &      & $R$ & 450 & 2.5 \nl
                                   &       &                        &      & $I$ & 600 & 2.5 \nl
                                   &       &                        &      & $H\alpha$ & 1800 & 2.5 \nl
            IRAS 11298+5313 (I283) & 03/92 & 3.6m CFHT-SILFID-Lick2 & 0.178 & $B$ & 600 & 0.9 \nl
                                   &       &                        &       & $V$ & 300 & 0.9 \nl
                                   &       &                        &       & $R$ & 300 & 0.9 \nl
                                   &       &                        &       & $H\alpha$\tablenotemark{3} & 1800 & 1.4\tablenotemark{4} \nl
                                   &       &                        &       & [\ion{O}{3}]$\tablenotemark{3}$ & 2400 & 1.4\tablenotemark{4} \nl
            IRAS 09453+5043 (I259) & 03/92 & 3.6m CFHT-SILFID-Lick2 & 0.178 & $V$ & 300 & 0.8\nl
                                   &       &                        &       & $R$ & 300 & 0.8 \nl
            IRAS 05238-4602 (I179) & 09/95 & 0.9m Dutch-Adapter-No33 & 0.443 & $R$ & 600 & 2.0 \nl
            IRAS 02366-3101 (I55)  & 08/92 & 1.54m Danish-Adapter-No17 & 0.298 & $B$ & 600 & 1.8 \nl
                                   &       &                          &       & $V$ & 300 & 1.8 \nl
                                   &       &                          &       & $R$ & 300 & 1.7 \nl
                                   &       &                          &       & $I$ & 300 & 1.8 \nl
                                   &       &                          &       & [\ion{O}{3}] & 2100 & 1.6 \nl
            IRAS 23254+0830 (I555) & 08/92 & 1.54m Danish-Adapter-No17 & 0.298 & $B$ & 420 & 1.6 \nl
                                   &       &                          &       & $V$ & 240 & 1.5 \nl
                                   &       &                          &       & $R$ & 240 & 1.7 \nl
                                   &       &                          &       & $I$ & 240 & 1.3 \nl
                                   &       &                          &       & $H\alpha$ & 1800 & 1.6 \nl
                                   &       &                          &       & [\ion{O}{3}] & 1800 & 1.5 \nl
            IRAS 00321-0019 (I9)   & 10/95 & 0.9m Dutch-Adapter-No33 & 0.443 & $V$ & 1200 & 1.4 \nl
                                   &       &                        &       & $R$ & 900  & 1.3 \nl
            IRAS 01346-0924 (I28)  & 10/95 & 0.9m Dutch-Adapter-No33 & 0.443 & $R$ & 900 & 1.7 \nl
            IRAS 02553-1642 (I65)  & 09/95 & 0.9m Dutch-Adapter-No33 & 0.443 & $B$ & 1050 & 2.3 \nl
                                   &       &                        &       & $R$ & 600  & 2.1 \nl
            IRAS 16382-0613 (I418) & 04/95 & 1.54m Danish-DFOSC-No17 & 0.51 & $I$ & 180 & 0.9 \nl
            IRAS 03278-4329 (I90)  & 08/95 & 1.54m Danish-DFOSC-No17 & 0.51 & $I$ & 600 & 1.9 \nl
                                   & 10/95 & 0.9m Dutch-Adapter-No33 & 0.443 & $R$ & 1050 & 1.6 \nl
            IRAS 04339-1028 (I139) & 09/95 & 0.9m Dutch-Adapter-No33 & 0.443 & $B$ & 420 & 1.9 \nl
                                   &       &                        &       & $R$ & 150 & 1.7 \nl
            IRAS 20208-5635 (I502) & 08/95 & 1.54m Danish-DFOSC-No17 & 0.51 & $I$ & 600 & 2.0 \nl
            IRAS 04229-2528 (I122) & 09/95 & 0.9m Dutch-Adapter-No33 & 0.443 & $R$ & 600 & 1.6 \nl
            IRAS 19254-7245 (I489) & 08/92 & 1.54m Danish-Adapter-No17 & 0.298 & $B$ & 900 & 1.8 \nl
                                   &       &                          &       & $V$ & 240 & 1.8 \nl
                                   &       &                          &       & $R$ & 270 & 1.8 \nl
                                   &       &                          &       & $I$ & 360 & 1.8 \nl
                                   &       &                          &       & $H\alpha$ & 1200 & 1.4 \nl
                                   &       &                          &       & [\ion{O}{3}] & 1200 & 1.3 \nl
            IRAS 08277-0242 (I245) & 03/92 & 3.6m CFHT-SILFID-Lick2 & 0.178 & $V$ & 300 & 0.7 \nl
                                   &       &                        &       & $R$ & 300 & 0.7 \nl
                                   & 04/95 & 1.54m Danish-DFOSC-No17 & 0.51 & $I$ & 420 & 1.9 \nl
            IRAS 03230-5800 (I84)  & 08/95 & 1.54m Danish-DFOSC-No17 & 0.51 & $V$ & 600 & 2.7 \nl
                                   &       &                        &      & $I$ & 600 & 3.1 \nl
            IRAS 23016+2221 (I547) & 08/92 & 1.54m Danish-Adapter-No17 & 0.298 & $B$ & 1050 & 1.6 \nl
                                   &       &                          &       & $V$ & 900  & 2.0 \nl
                                   &       &                          &       & $R$ & 240  & 1.9 \nl
                                   &       &                          &       & $I$ & 240  & 2.0 \nl
                                   &       &                          &       & $H\alpha$ & 1500  & 2.2 \nl
                                   &       &                          &       & [\ion{O}{3}] & 1925 & 2.2 \nl
            IRAS 06563-6529 (I213) & 04/95 & 1.54m Danish-DFOSC-No17 & 0.51 & $I$ & 420 & 1.3 \nl
\enddata
\tablenotetext{1} {In most cases, the integration times are the average of several exposures}
\tablenotetext{2} {The ``seeing'' is measured on the field stars (FWHM)}
\tablenotetext{3} {Reconstructed emission line images, from Integral Field Spectroscopy with the fiber spectrograph SILFID (Chatzichristou and Vanderriest 1995 and Chatzichristou et al. 1998)}
\tablenotetext{4} {The spatial sampling of the fiber spectrograph (1.4 or 0.7 arcsec/fiber)}

\end{deluxetable}

\clearpage